\documentclass[aps,prl,twocolumn,superscriptaddress]{revtex4}

\def\av#1{\left\langle#1\right\rangle}

\usepackage{latexsym}
\usepackage{amstext,amssymb,amsfonts}
\usepackage{epsfig}

\begin{document}


\title{Priority diffusion model in lattices and complex networks}


\author{Michalis Maragakis}
\thanks{These authors contributed equally to this work}
\affiliation{Department of Physics, University of Thessaloniki, 54124 Thessaloniki, Greece}
\author{Shai Carmi}
\thanks{These authors contributed equally to this work}
\affiliation{Minerva Center \& Department of Physics, Bar-Ilan University, Ramat Gan 52900, Israel}
\author{Daniel ben-Avraham}
\affiliation{Department of Physics, Clarkson University, Potsdam NY 13699-5820, USA}
\author{Shlomo Havlin}
\affiliation{Minerva Center \& Department of Physics, Bar-Ilan University, Ramat Gan 52900, Israel}
\author{Panos Argyrakis}
\affiliation{Department of Physics, University of Thessaloniki, 54124 Thessaloniki, Greece}


\date{\today}

\begin{abstract}

We introduce a model for diffusion of two classes of particles ($A$ and $B$) with priority:
where both species are present in the same site the motion of $A$'s takes precedence over that of $B$'s.
This describes realistic situations in wireless and communication networks.
In regular lattices the diffusion of the two species is normal
but the $B$ particles are significantly slower, due to the presence of the $A$ particles.
From the fraction of sites where the $B$ particles can move freely, which we compute analytically,
we derive the diffusion coefficients of the two species.
In heterogeneous networks the fraction of sites where $B$ is free
decreases exponentially with the degree of the sites. This, coupled with accumulation
of particles in high-degree nodes leads to trapping of the low
priority particles in scale-free networks.

\end{abstract}


\maketitle


Diffusion, or the random motion of particles is a most basic
mechanism underlying numerous phenomena in nature and technology.
While diffusion in periodic regular lattices is rather simple, it is
significantly richer in disordered media and complex networks, or
when the particles interact with one another. In most studies the
particles interact through the excluded volume effect, combined with
a potential (as in `lattice gas' models), or through some kind of
reaction or transformation of the particles (e.g.
\cite{Weiss_book,DbA_book,VanKampen_book,Redner_book}).

In this paper we introduce a two-species priority diffusion model
(PDM): the two species, $A$ and $B$, diffuse independently, but
where both species coexist only the high priority particles, $A$,
are allowed to move. This problem has several important
applications. A frequent case in communication networks is that data
packets traverse the networks in a random fashion (e.g. in wireless
sensor networks \cite{RW_sensor1,RW_sensor2}, ad-hoc networks
\cite{RW_adhoc1,RW_adhoc2} and peer-to-peer networks \cite{RW_P2P}).
Routers in communication networks handle both high and low priority
information packets, such as, for example, in typical multimedia
applications. The low priority packets are sent out only after all
high priority packets have been sent \cite{KR_book,Tanenbaum_book},
just as in our model.


We solve the priority diffusion model (PDM) analytically for
lattices and networks. In lattices and regular graphs both species
diffuse in the usual fashion, but the low priority $B$'s diffuse
slower than the $A$'s. In heterogeneous scale-free networks the
$B$'s get mired in the high degree nodes, effectively arresting
their progress. We confirm these conclusions through large-scale
computer simulations.

The $A$ and $B$ particles, when selected for motion, hop to one of
the nearest neighbor sites, with equal probability. We have
investigated two selection protocols. In the \emph{site} protocol a
site is selected at random: if it contains both $A$ and $B$
particles, a high-priority $A$ particle moves out of the site. A
particle of type $B$ moves only if there are no $A$'s on the site.
If the site is empty, a new choice is made. In the \emph{particle}
protocol a particle is randomly selected: if the particle is an $A$
it then hops out. A selected $B$ hops only if there are no $A$
particles on its site \cite{other_particle_protocol}. The site
protocol describes the case where a selected router sends out
packets of information, whereas in the particle protocol agents move
independently, describing perhaps the commuting of various
individuals. Note that these protocols belong to the general
framework of zero-range process with two species of particles (see
e.g. \cite{ZeroRange,ASEP} for factorized steady-state solutions).

We note that the $A$ particles essentially move freely (in both
protocols), regardless of the $B$'s. The $B$'s, on the other hand,
can move only in those sites that are empty of $A$'s. We begin by
considering the number of such sites.
We later relate this property to diffusion coefficients of the particles
under the priority constraints.

We look first at lattices, or \emph{regular} graphs, where each site
has exactly $z$ nearest neighbors. The number of sites
$N\rightarrow\infty$ and for now we focus on a single species,
denoting its particle density by $\rho$. Let $f_j$ be the average
(equilibrium) fraction of sites that contain $j$ particles. Consider
a Markov chain  process whose states $\{0,1,2,...\}$ are the number
of particles in a given site.  The $\{f_j\}_{j=0,1,2,...}$ are the
stationary probabilities of the chain.

\begin{figure}[t]
\hskip -0.1cm
\epsfig{file=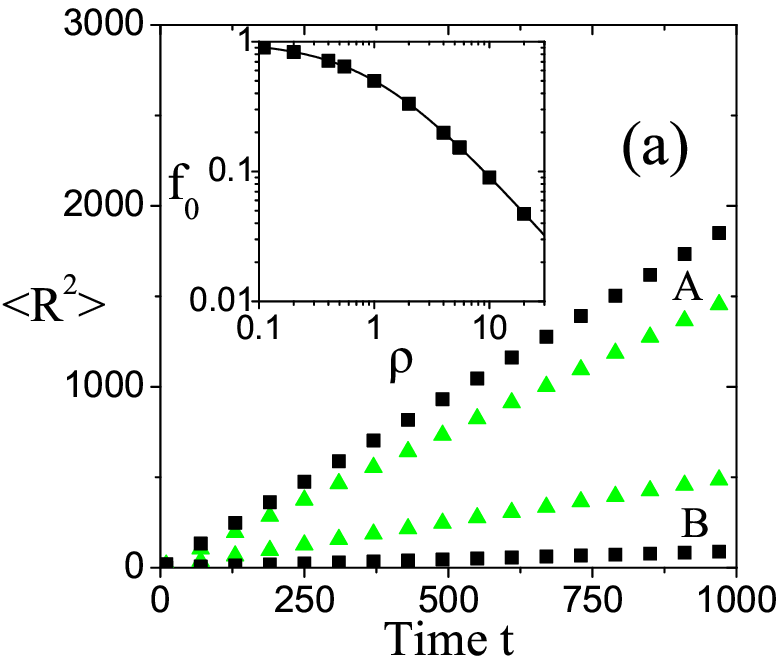,height=4.5cm,width=4.5cm}
\hskip -0.4cm
\epsfig{file=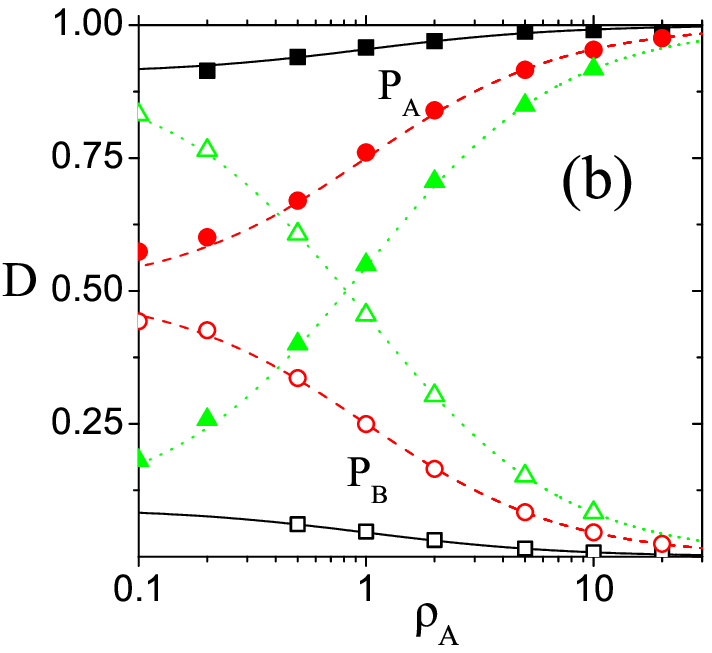,height=4.5cm,width=4.5cm}
\caption{(Color online) PDM in lattices (site protocol). Typical
system size is $10^4$ sites and time is $10^3$ steps. (a)~Mean
square displacement of a particle $\av{R^2}$ (averaged over all
particles and realizations) as a function of time. In one time step
each particle moves on average once. Black squares are for densities
$\rho_A=\rho_B=10$, green triangles for $\rho_A=\rho_B=1$. Inset:
$f_0$ as a function of density. The solid curve represents  Eq.
(\ref{empty_site}), while symbols denote simulation results.
(b)~Diffusion coefficients, $D$ ($P_A$ full symbols, $P_B$ empty
symbols), as a function of $\rho_A$. Black squares are for
$\rho_A=10\rho_B$, red circles for $\rho_A=\rho_B$ and green
triangles for $\rho_A=0.1\rho_B$. The continuous ($P_A$) and dotted
($P_B$) curves represent Eq.~(\ref{P_A}).
}
\label{Fig1}
\end{figure}

For the site protocol, the transition probabilities are:
\begin{equation}
\label{eq1}
P_{j,j-1} = \frac{1}{N}\;;\qquad P_{j,j+1} =
\frac{1-f_0}{N}\;.
\end{equation}
$P_{j,j} = 1-P_{j,j-1}-P_{j,j+1}$ and all other transitions cannot
occur. Indeed, for a site to lose a particle it needs to be
selected, with probability $\frac{1}{N}$. To gain a particle, a
non-empty, one of its $z$ neighbors must be chosen --- with
probability $(1-f_0)\frac{z}{N}$ --- and this neighbor must send the
particle into the original site, with probability $\frac{1}{z}$.
Note that the final result is independent of the coordination number
$z$. The stationary state satisfies
\begin{equation}
\label{stationary}
f_j=\sum_{i=0}^{\infty}f_iP_{ij}\;,
\end{equation}
or, in view of~(\ref{eq1}),
\begin{equation}
\label{site_lattice}
f_{j-1}(1-f_0)+f_{j+1}=f_j+f_j(1-f_0)\;,
\end{equation}
with the boundary condition $f_1=(1-f_0)f_0$. This has the solution
$f_j = f_0(1-f_0)^j$. Imposing particle conservation
$\sum_{j=0}^{\infty}jf_j=\rho$, we finally obtain:
\begin{equation}
\label{empty_site}
f_0^{(\mbox{site})} = \frac{1}{1+\rho}\;.
\end{equation}
This agrees nicely with our simulation results (Fig.~\ref{Fig1}(a), inset).

For the particle protocol the transition
probabilities are:
\begin{equation}
\label{transition_particle}
P_{j,j-1} = \frac{j}{N\rho}\;;\qquad P_{j,j+1} =
\frac{1}{N}\;,
\end{equation}
$P_{j,j} = 1-P_{j,j-1}-P_{j,j+1}$ and all other transitions are excluded.
Indeed, for a site to lose a particle
one of its $j$ particles (out of the total $N\rho$) needs to be selected.
To gain a particle, one of the $z\rho$ particles that reside, on average,
in the neighboring sites has to be chosen,
and then hop to the original site (with probability
$\frac{1}{z}$). Once again, the result is independent of $z$.
This time the boundary condition is $f_1=\rho f_0$, leading to
$f_j=f_0\frac{\rho^j}{j!}$. Imposing the normalization
condition $\sum_{j=0}^{\infty}f_j=1$, one finally finds
\begin{equation}
\label{empty_particle}
f_0^{(\mbox{particle})} = e^{-\rho}\;.
\end{equation}
In other words, the $\{f_j\}$ are Poisson-distributed, with average $\rho$. 

We now employ these results for the analysis of priority diffusion,
when \emph{both} species are involved.
In regular graphs both species diffuse as in the single-species case, but
due to the priority constraints the total time available is split unevenly between the $A$'s and $B$'s.
For example, in lattices diffusion is normal,
$\av{R^2}=Dt$, as confirmed by simulations (Fig.~\ref{Fig1}(a)),
but with a smaller diffusion coefficient for the hindered $B$'s.

Denote by $P_A$ (resp. $P_B$) the probability that a moving particle
is an $A$ ($B$). In the site protocol, a particle will surely move
if we choose a non-empty site (contains $A$, $B$ or both), which
happens with probability $1-1/(1+\rho_A+\rho_B)$ (since the
particles behave as a single, non-interacting species, if one
ignores their labelling, and thus eq. (\ref{empty_site}) can be
applied with $\rho=\rho_A+\rho_B$). $A$ moves if the selected site
contains any number of $A$'s, which happens with probability
$\rho_A/(1+\rho_A)$, again, from (\ref{empty_site}). Therefore,
$P_A=\frac{\rho_A}{1+\rho_A}/(1-\frac{1}{1+\rho_A+\rho_B})$, or:
\begin{equation}
\label{P_A} P_A = \frac{\rho_A(1+\rho_S)}{(1+\rho_A)\rho_S}\,,\quad
P_B = \frac{\rho_B}{(1+\rho_A)\rho_S}\,,
\end{equation}
where $\rho_S\equiv\rho_A+\rho_B$ and we have used $P_B=1-P_A$ for
the second relation. For a single particle in a lattice,
$\av{R^2}=t$. Here we have $\av{R_A^2}=P_At\equiv D_At$ and thus the
diffusion coefficient $D_A$ equals $P_A$ (similarly, $D_B=P_B$).
Simulations shown in Fig.~\ref{Fig1}b confirm our predictions
(Eq.~(\ref{P_A})).

For the particle protocol, denote the ratio of free $B$ particles
(that do not share a site with $A$'s) to all $B$ particles by $r$.
Any particle will surely move except for the case when a non-free
$B$ particle was chosen (which happens with probability
$\rho_B/\rho_S\cdot (1-r)$). $B$ moves whenever a free $B$ is
chosen, with probability $\rho_B/\rho_S\cdot r$. Therefore we find:
\begin{equation}
\label{P_B} P_B = \frac{r\rho_B/\rho_S}{1-(1-r)\rho_B/\rho_S},
\end{equation}
and $P_A=1-P_B$.

Had the density of $B$'s been independent of the $A$'s
then $r$ would simply be the fraction of sites empty of $A$, or $r=e^{-\rho_A}$.
However, due to the priority constraints $B$'s tend to stick with the
$A$'s, so that $r \lesssim e^{-\rho_A}$.
The ratio $r$ can be obtained analytically for low densities,
if we assume that a single site cannot contain more than one $A$ or
$B$. We use again a Markov chain formulation,
but now with just four possible states to each site: $\{\phi,A,B,AB\}$
(state $A$ corresponds to a site having one $A$ particle, and similarly
for the other states).
We write the transition probabilities
as before, to first order in the densities:
\begin{eqnarray}
\nonumber && P_{\phi,A} = \frac{\rho_A}{N\rho_S} \;;\quad
P_{\phi,B} = \frac{\rho_B}{N\rho_S} \;;
\\ \nonumber &&
P_{A,\phi} = \frac{1}{N\rho_S} \;;\quad
P_{A,AB} = \frac{\rho_B}{N\rho_S}  \;;
\\ \nonumber &&
P_{B,\phi} = \frac{1}{N\rho_S} \;;\quad
P_{B,AB} = \frac{\rho_A}{N\rho_S} \;;
\\ &&
P_{AB,B} = \frac{1}{N\rho_S}\;.
\end{eqnarray}
Unindicated transition probabilities are zero, and the diagonal
accounts for normalization $P_{x,x}=1-\sum_{y \ne x}P_{x,y}$.
The justification is similar to that of Eq.~(\ref{transition_particle}).
For a site to lose a particle, this particle needs to be
chosen out of a total of $N\rho_S$ particles.
For a site to gain an $A$, one of the $z\rho_A$
particles that reside, on average, in the neighboring sites
has to be chosen (out of $N\rho_S$),
and then sent to the target site, with probability $\frac{1}{z}$
(and likewise for gaining a $B$).
The priority constraint is taken into account by forbidding the transition $AB \rightarrow A$.

\begin{figure}[t]
\hskip -0.1cm
\epsfig{file=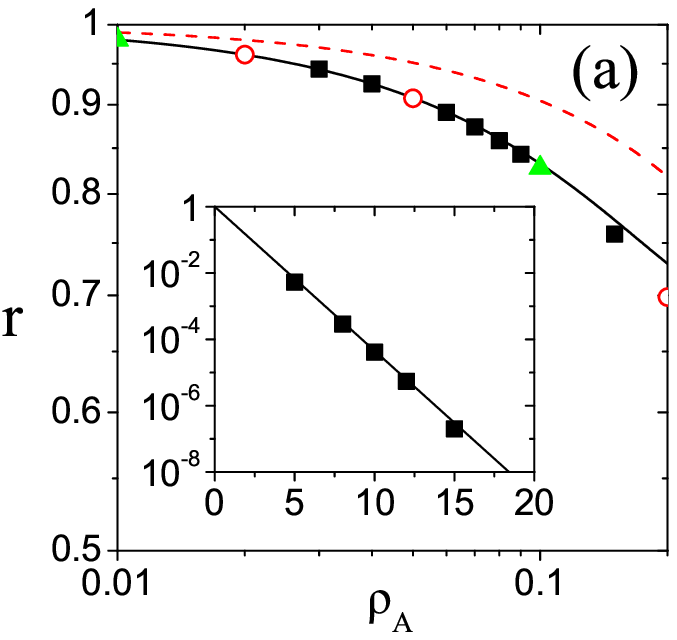,height=4.5cm,width=4.5cm}
\hskip -0.4cm
\epsfig{file=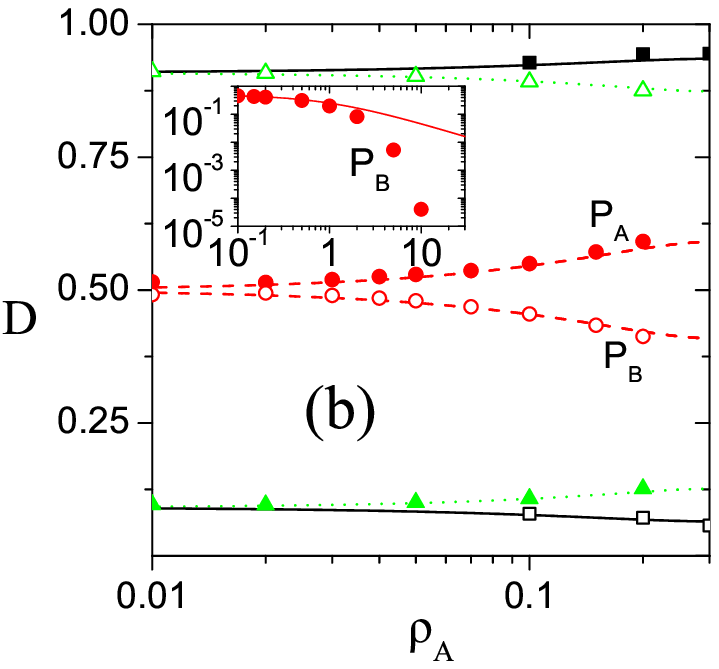,height=4.5cm,width=4.5cm}
\caption{(Color online) PDM in lattices (particle protocol).
Simulation parameters are same as in Fig.~\ref{Fig1}. (a)~Fraction
$r$ of free $B$'s as a function of $\rho_A$ for the cases
$\rho_A=\rho_B$ (black squares), $\rho_A=0.2\rho_B$ (empty red
circles), and $\rho_A=5\rho_B$ (green triangles), compared to the
result of Eq.~(\ref{r_second_order}) (solid line). $r$ is smaller
than $e^{-\rho_A}$ (dotted red line) since $B$ particles tend to
stick with the $A$'s. Inset: Large $\rho_A$ regime, where $r$ tends
to $e^{-\rho_A}$ ($\rho_B=\rho_A$). (b)~Diffusion coefficients
($P_A$ full symbols, $P_B$ empty symbols) as a function of $\rho_A$
for the cases $\rho_A=10\rho_B$ (black squares), $\rho_A=\rho_B$
(red circles), and $\rho_A=0.1\rho_B$ (green triangles). The
continuous ($P_A$) and dotted ($P_B$) curves represent
Eqs.~(\ref{P_B}) and (\ref{r_second_order}). Inset shows $P_B$ (red
circles) for large $\rho_A$. The $B$ particles diffuse much slower
here compared to the site protocol (continuous red line) due to the
exponential decrease in the number of empty sites. } \label{Fig2}
\end{figure}

From the stationary probabilities of the chain (Eq.~(\ref{stationary}))
we derive $r$ to first order:
\begin{equation}
\label{r_first_order}
r = \frac{f_B}{f_B+f_{AB}} = 1-2\rho_A+{\cal O}(\rho^2)
\end{equation}
($\rho$ stands for either $\rho_A$ or $\rho_B$).
To obtain the next order, allowed states can have two particles
of each type, and we take into account that when a $B$ is chosen
it actually hops only with probability $r$ (using its first-order expression, Eq.~(\ref{r_first_order})).
We thus find
\begin{equation}
\label{r_second_order}
r = 1-2\rho_A+\frac{13}{4}\rho_A^2 + {\cal O}(\rho^3)\;.
\end{equation}
Surprisingly, $r$ does not depend on $\rho_B$, at least to second order.
In fact, our simulations suggest that $r$ is independent of $\rho_B$
for all densities (Fig.~\ref{Fig2}(a)). For large $\rho_A$ we have $r \rightarrow e^{-\rho_A}$,
since free $B$'s become extremely rare.
The diffusion coefficients obtained on substituting Eq.~(\ref{r_second_order}) in (\ref{P_B})
compare quite nicely with simulations (Fig.~\ref{Fig2}(b)).

We now turn to heterogeneous networks, where the degree $k$ varies from site to site.
We focus on two explicit models:
Erd\H{o}s-R\'{e}nyi (ER) random graphs \cite{ER,Bollobas},
where the degrees of the nodes are narrowly (Poisson) distributed, and scale-free (SF) networks,
recently discovered to best describe a wide variety of natural and man-made systems,
and in particular many communication networks such as the Internet.
In SF nets the degree distribution is broad, characterized by a power-law tail
$P(k)\sim k^{-\gamma}$, when usually $2<\gamma<3$ \cite{BA_review,PV_book,DM_book,Newman_review}.

\begin{figure}[t]
\hskip -0.1cm
\epsfig{file=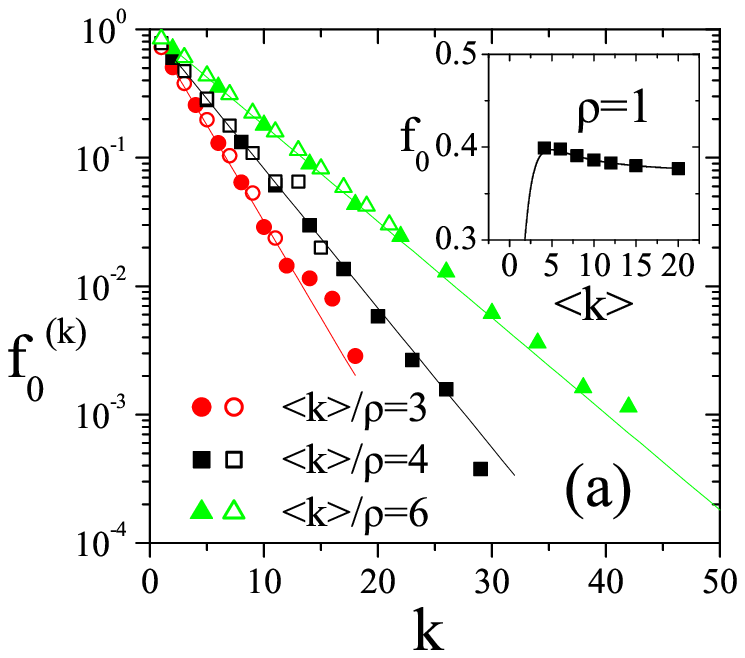,height=4.5cm,width=4.5cm}
\hskip -0.4cm
\epsfig{file=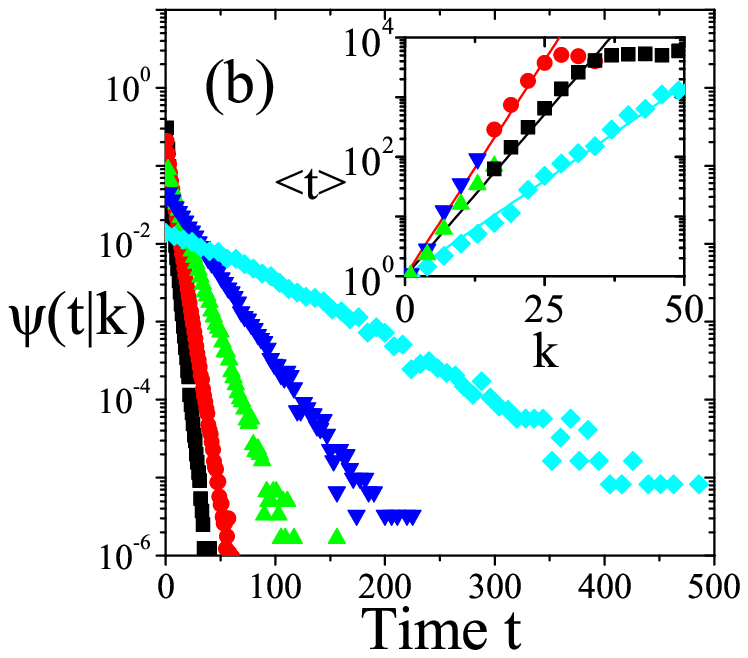,height=4.5cm,width=4.5cm}
\caption{(Color online) PDM in networks. Typical system size for
both ER and SF is $10^4$ nodes, Internet is 21955 nodes and time is
$10^4$ MCS. SF networks were generated using the Molloy-Reed
algorithm \cite{MR}. (a)~Fraction of empty sites in ER (empty
symbols) and SF (full symbols) networks as a function of site
degree. The curves follow Eq.~(\ref{empty_k}). Inset: Fraction of
empty sites for ER vs.~$\av{k}$ for fixed density $\rho=1$. The
continuous curve represents Eq.~(\ref{empty_er}). (b)~Distribution
of waiting times for $B$ particles for nodes of various degrees
($k=2,4,7,10,15$, left to right) in SF networks ($\gamma=3,m=2$).
Inset: $\av{t}$ as a function of k. Shown are: SF and ER networks
with $\av{k}=4$ (black squares and green triangles), and $\av{k}=3$
(red circles and blue triangles), and the Internet \cite{DIMES}
(cyan diamonds). Solid curves represent the theoretical
$\av{t(k)}=\exp(\rho_A k/\av{k})$. Plateau for large $k$ results
from finite simulation time.} \label{Fig3}
\end{figure}

We analyze the particle protocol only --- the site protocol yields
similar results, as we confirmed through computer simulations. We
wish to find the fraction of empty sites \emph{of degree} $k$,
$f_0^{(k)}$. As before, consider a network with only one particle
species and define a Markov chain on the states $\{0,1,2...\}$ for
the number of particles in a given site of degree $k$. The
stationary probabilities are $f_j^{(k)}$. It can be shown that the
chain has the transition probabilities:
\begin{equation}
P_{j,j-1} = \frac{j}{N\rho}\;;\quad P_{j,j+1} =
\frac{k}{\av{k}}\frac{1}{N}\;,
\end{equation}
$P_{j,j} = 1-P_{j,j-1}-P_{j,j+1}$ and all other probabilities are
zero. Intuitively, this is the same as
Eq.~(\ref{transition_particle}), except that here a site may gain a
particle from any of its $k$ neighbors, and the neighbor delivers
the particle to the target site with probability $\frac{1}{\av{k}}$
($\av{k}$ is the average degree of the net). Solving for the
stationary probabilities while keeping in mind that $\sum_j
f_j^{(k)}=1$ one finds
\begin{equation}
\label{empty_k}
f_j^{(k)} = f_0^{(k)}\frac{\left(\frac{\rho k}{\av{k}}\right)^j}{j!}\;;\quad
f_0^{(k)} = \exp\left({-\frac{\rho k}{\av{k}}}\right)\;.
\end{equation}
Note that for regular graphs, when all sites have the same degree,
this reduces to Eq.~(\ref{empty_particle}), $f_0=e^{-\rho}$. Also,
the total number of particles in a site of degree $k$ is thus
$\sum_{j=0}^{\infty}jf_j^{(k)}=\frac{\rho k}{\av{k}}\propto k$, as
is well known for random walks on networks~\cite{Rieger}. For the
full network, $f_0=\sum_{k=1}^{\infty}f_0^{(k)}P(k)$ \cite{k_0}. In
particular, for ER graphs,
\begin{equation}
\label{empty_er} f_0 = \sum_{k=1}^{\infty}e^{-\frac{\rho
k}{\av{k}}}e^{-\av{k}}\frac{\av{k}^k}{k!} =
e^{-\av{k}} (e^{\av{k}e^{-\frac{\rho}{\av{k}}}}-1)\,.
\end{equation}
The agreement between these predictions and simulations is shown in Fig.~\ref{Fig3}(a).

Consider now priority diffusion. Define that in one time step each
particle has on average one moving attempt. On average, at every
time step a $B$ particle (in a node of degree $k$) has a probability
$\exp(-\rho_A k/\av{k})$ to be able to jump out
(Eq.~(\ref{empty_k})). This results in a distribution of waiting
times (for a $B$ particle)
\begin{equation}
\psi_k(t) = \frac{1}{\tau}e^{-t/\tau}\;,
\end{equation}
where $\tau\equiv\av{t(k)}=\exp(\rho_A k/\av{k})$
is simply the inverse of the probability for the site to be empty of $A$'s (Fig.~\ref{Fig3}(b)).

The exponentially long waiting time (in the degree $k$) means that
in heterogeneous networks such as scale-free nets --- where the
degrees may span several orders of magnitude --- the $B$ particles
get mired in the hubs (high degree nodes). The problem is
exacerbated by the fact that the $B$ particles are drawn to the hubs
even in the absence of $A$'s: the presence of $A$'s only amplifies
this tendency, because of the positive correlation between the
concentrations of the two species. Thus, while the concentration of
the $A$'s is proportional to $k$, it can be shown that the
concentration of the $B$'s is proportional to $k\exp(\rho_A
k/\av{k})$. In large scale-free nets the $B$'s collect at the hubs,
which results in extremely long waiting times and practical halting
of particles diffusion.

In conclusion, we have introduced and analyzed a model of priority diffusion, involving two species,
where the particles of the preferred species always move ahead of the other.
In regular graphs and lattices diffusion of each of the species is similar to that of a single
(non-interacting) species on the same substrate, but the overall diffusion rate of the preferred
species is significantly faster: we have provided exact expressions for the rate ratios in this case.
In heterogeneous nets, we have shown that the slow-species particles tend to be collected in the
high degree nodes, where their waiting time for clearing the site increases exponentially with the
degree of the site.
In scale-free nets where the degrees span several orders of magnitude the slow species
progress is effectively arrested.

Our calculations for heterogeneous networks are mean-field in
character, in that we ignore possible correlations between the
degrees of neighboring nodes. This assumption is valid, though, for
ER random graphs and for maximally random networks, such as
Molloy-Reed SF nets \cite{MR}. The possible effect of degree-degree
correlations remains a subject for future research.


\begin{acknowledgments}
We thank L.K. Gallos and H. Rozenfeld for useful discussions.
Financial support from the NSF, the Israel Science Foundation, the
Israel Center for Complexity Science, the European NEST project
DYSONET, and GSRT project PENED 03ED840, is gratefully acknowledged.
S.C. is supported by the Adams Fellowship Program of the Israel
Academy of Sciences and Humanities
\end{acknowledgments}

\bibliography{mcbha}

\begin{thebibliography}{24}
\bibitem[{oth()}]{other_particle_protocol}
\bibinfo{note}{We have also studied a third protocol where an $A$ moves when a
  $B$ is selected in a site containing $A$'s. A full report will be published
  elsewhere.}
\bibitem[{k_0()}]{k_0}
\bibinfo{note}{If all sites are initally occupied, isolated sites ($k=0$) are
  never empty, and thus they are excluded from the sum.}
\end{thebibliography}

\end{document}